\documentclass[twocolumn,twoside,slac_two]{revtex4}
\usepackage{graphicx}
\usepackage{fancyhdr}
\begin{document}

\title{GRB jets inside and outside the star: precursors and
cosmological implications}

%

\author{Davide Lazzati, Mitchell C. Begelman} \affiliation{JILA, 440
UCB, University of Colorado, Boulder, CO 80309-0440, USA}

\author{Giancarlo Ghirlanda, Gabriele Ghisellini, Claudio Firmani}
\affiliation{Osservatorio Astronomico di Brera, via E. Bianchi 46,
I-23807 Merate (LC), Italy}

\begin{abstract}
After many years of speculation, recent observations have confirmed
the association of gamma-ray bursts with core-collapse supernova
explosions from massive stars. This association carries with it
important consequences. The burst relativistic jet has to propagate
through the cold dense stellar material before it reaches the
transparency radius and the burst photons are produced. This
propagation is likely to affect the initial properties of the jet,
shaping it and changing its energy composition. The variability
injected at the base of the jet is also likely to be erased by the
jet-star interaction. Despite this, GRBs seem to have remarkably
predictable properties once the radiative phase sets in, as
emphasized by the recent discovery of several tight correlation
between spectral, geometric and energetic properties of the jet.  In
this contribution we discuss the jet interaction with the star,
emphasizing its time-dependent properties and the resulting energy
distribution. We finally emphasize the surprising predictability of
jet and radiation properties outside the star and underline its
implication for standardizing the GRB candle.
\end{abstract}

\maketitle

\thispagestyle{fancy}

\section{Introduction}

After many years of speculation~\cite{ref:wos93,ref:pac98}, the
association of (at least some) long gamma-ray bursts (GRBs) with
supernova (SN) explosions has been established. The first association
was between GRB~980425 and SN1998bw~\cite{ref:gal98}. It was a
problematic and highly debated association since it implied a highly
peculiar nature for GRB~980425: an event with a normal fluence but a
distance 100 times smaller then other GRBs, and therefore some four
orders of magnitude fainter. In addition, GRB~980425 did not have a
normal afterglow in any band. Nevertheless, the peculiar nature of the
SN explosion and the temporal and spatial coincidence argued strongly
for the association. After GRB~980425/SN1998bw, circumstantial evidence
for more associations was claimed for higher redshift events, based on
the appearance of a red bump in the late afterglow
light curve~\cite{ref:zeh04}.

More recently two additional events have strengthened the association,
including - at least in one case - one GRB that has all the
characteristics of classical cosmological events. GRB~030329 was the
brightest GRB detected by HETE-2~\cite{ref:van04}. It had a small
redshift $z=0.168$ and multi-epoch afterglow spectroscopy revealed the
emergence of a SN spectrum approximately 15 days after the burst
explosion~\cite{ref:sta03,ref:hjo03,ref:mat03}. The spectrum and the
overall supernova properties were remarkably similar to those of
SN1998bw. Finally, GRB~031203, detected by INTEGRAL, looks like a twin
to GRB~980425/SN1998bw~\cite{ref:mal04}.

The association bears new questions and riddles with it. First we know
that the $\gamma$-ray photons are produced by ultra-relativistic
material. The progenitor star is however dense and massive, and any
outflow collecting matter geometrically from the star will be
slowed-down to sub-relativistic speed very effectively. The entrained
mass can be at most a small fraction of a solar mass
$M_0\le5.5\times10^{-6}\,E_{51}\,\Gamma_2^{-1}\,M_\odot$. Simulations
show that such a small contamination is possible (see below). The jet
that reaches the surface of the star is then shaped by the interaction
with the star itself. One would expect that, depending on the type of
stellar progenitor, jet initial conditions and energetics, a great
diversity of jets would emerge. After all type Ibc and type II
supernovae show a remarkable diversity.

Even though at first sight GRB observations are characterized by a
huge diversity, remarkable correlations have been discovered among jet
properties. The first such correlation was discovered between the jet
$\gamma$-ray energy output (not corrected for the beaming) and the
break time in the afterglow
light curve~\cite{ref:fra01,ref:pan01}. This correlation can be
understood in terms of jets with the same total energy but beamed
into different opening angles~\cite{ref:fra01}, or in terms of
structured jets with an energy distribution such that
$dE/d\Omega\propto\theta^{-2}$~\cite{ref:ros02}.

A further correlation was discovered between the typical frequency of
photons in the prompt emission (the peak of the $\nu\,F(\nu)$
spectrum) and the isotropic equivalent energy output in the prompt
phase~\cite{ref:ama02}. According to this correlation the brighter the
burst, the larger the typical frequency of photons. Finally, the
tightest correlation of all was recently discovered between the
typical photon energy and the beaming corrected $\gamma$-ray
output~\cite{ref:ghi04} (under the assumption of a uniform jet).  All
these correlations, and especially those involving the photon
frequency, have not been understood so far.

In this contribution we discuss the jet propagation inside the
progenitor star~\cite{ref:laz05}. We consider in particular the effect
of the diminishing collimating power of the star, which is blown apart
by the energy dissipated by the jet as it opens its way out of the
progenitor. We show that this creates time dependent effects that may
be responsible for the diversity in the $\gamma$-ray phase of the
burst. Such effects, however, occur on short timescales, and therefore
are not relevant in the afterglow phase, which is therefore more
standardized. Finally, we discuss the opportunity offered by the
Ghirlanda correlation~\cite{ref:ghi04} to use GRBs as standard candles
to constrain the expansion history of the universe at redshifts much
larger than the range sampled by type Ia SNe~\cite{ref:ghi04a}.

\section{The star-cocoon-jet interactions}

Consider a jet generated at the core of a massive star (at a radius
$R_0$), characterized by a luminosity $L_j$, an initial Lorentz factor
$\Gamma_0\gtrsim1$, a dimensionless entropy $\eta=L_j/(\dot{M}\,c^2)$
(where $\dot{M}$ is the rest mass ejection rate) and an initial
opening angle $\theta_0$ (or, alternatively, a stagnation pressure
$p_{\rm{stag}}=L_j\,\Gamma_0^2/(4\pi\,c\,R_0^2\,\theta_0^2)$). The jet
attempts to propagate through the stellar material. As soon as the
head of the jet reaches supersonic velocity, a double shock structure
develops at the head of the jet. The speed of the head of the jet (the
contact discontinuity) can be computed by balancing the pressure of
the jet material with that of the cold stellar matter. Pressure
balance reads~\cite{ref:mar94,ref:matz03}:
\begin{equation}
\rho_j\,h_j(\Gamma\beta)_{jh}^2\,c^2+p_j = \rho_\star\,h_\star
(\Gamma\beta)_h^2\,c^2+p_\star
\end{equation}
where $\rho$ is the density, $h$ the enthalpy and $p$ the
pressure. Subscripts are $_j$ for jet material, $_{jh}$ for the jet
material relative to the head of the jet, $_h$ for the jet head
relative to the star and $_\star$ for the stellar material. This
equation, if we neglect the pressure terms on both sides, can be
solved to give the motion of the jet head in the stellar
material~\cite{ref:mar94}:
\begin{equation}
\beta_h=\frac{\sqrt{\zeta\Gamma_j^2}}{1+\sqrt{\zeta\Gamma_j^2}}\,\beta_j
\end{equation}
where $\zeta=\rho_j\,h_j/\rho_\star$.

The actual structure of the head of the jet is made by a contact
discontinuity that separates jet material from stellar material, two
thin regions of shocked material (shocked jet and shocked stellar
material) and two shocks (a forward bow shock propagating into the
star and a reverse shock into the jet). The shocked material is
over-pressured and flows to the sides of the jet, feeding a cocoon.

The cocoon material is over-pressured with respect to both the star and
the un-shocked jet. It therefore expands, driving a shock into the cold
stellar material and collimating the jet until pressure balance is
reached. We assume an adiabatic flow, since the jet is optically thick
inside the star. As a consequence, the Lorentz factor scales as
$\Gamma_j\propto\Sigma_j^{1/2}$, where $\Sigma_j$ is the area of the
jet. This relation, which is valid only if dissipation is negligible,
allows us to compute the jet pressure as a function only of $\Sigma_j$
and to derive consequently the Lorentz factor. The cocoon evolution
can be computed by applying the first law of thermodynamics:
\begin{equation}
d(\epsilon_c\,V_c)=dQ-p_c\,dV_c
\end{equation}
where $\epsilon_c$ is the cocoon energy density and $V_c$ is the
volume of the cocoon. We assume that the cocoon pressure is uniform,
so that $p_c$ only depends on time. The term $dQ=L_j(1-\beta_h)dt$ is
the energy injected in the form of shocked jet material. The evolution
of the cocoon volume $V_c$ reads:
\begin{equation}
\frac{dV_c}{dt}=2\pi\int_{R_0}^{R_h}R_\perp(r,t)\,v_{sh}(r,t)\,dr
\end{equation}
where $R_\perp$ is the transverse size of the cocoon (which is a
function of radius and time) and $v_{sh}$ is the speed of the shock
driven by the cocoon pressure into the stellar material. It can be
computed under the Kompaneets approximation~\cite{ref:kom60} by
balancing the cocoon pressure against the ram pressure exerted by the
cold stellar material on the expanding cocoon:
$v_{sh}=\sqrt{\epsilon_c/3\,\rho_\star}$.

The above set of equations form a solvable system of differential
equations. An analytic solution is not possible, especially if the
time-dependent evolution of the cocoon needs to be taken into account.

\begin{figure*}
\includegraphics[width=.9\textwidth]{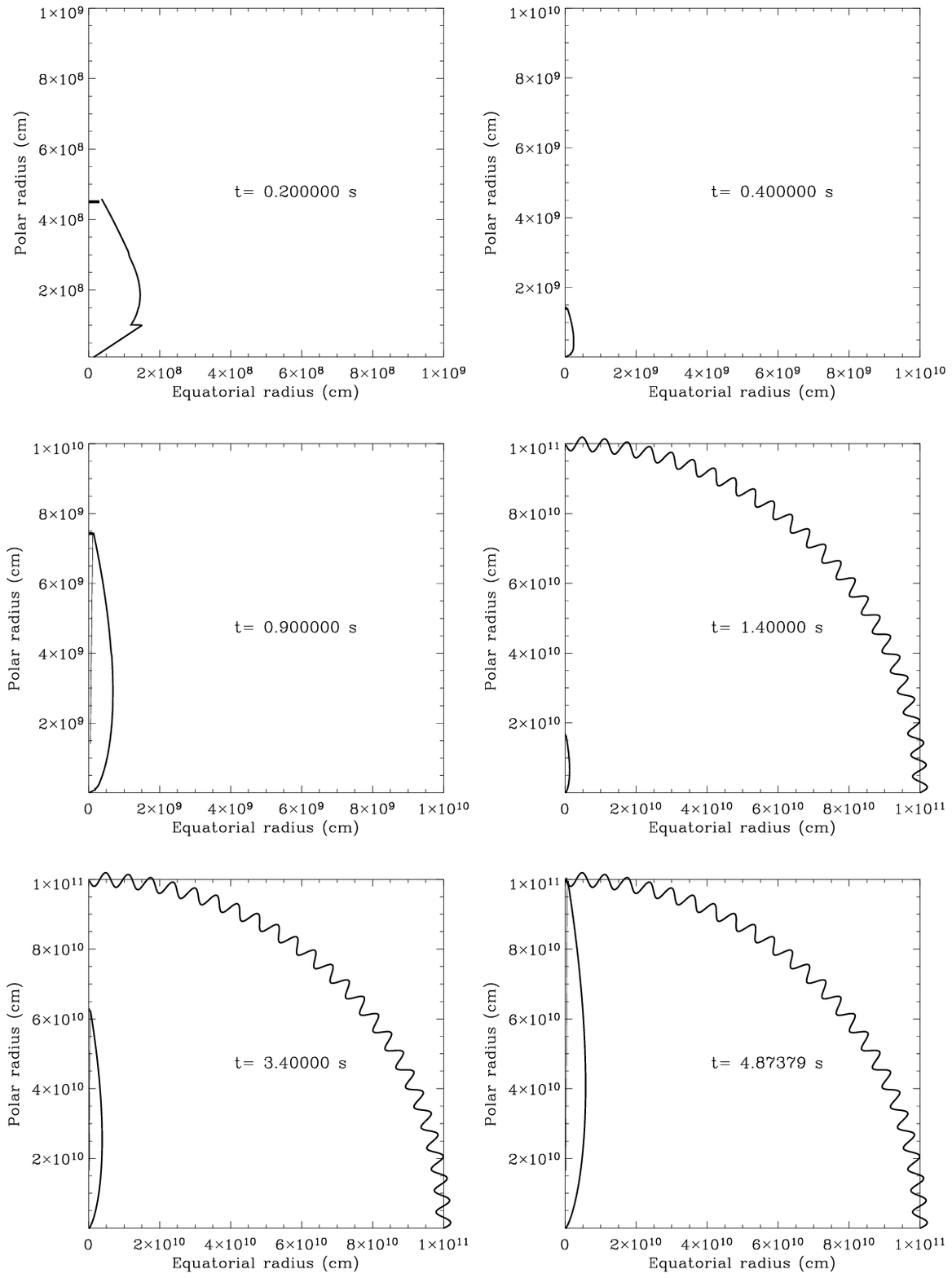}
\caption{{Stills from a movie showing the jet and cocoon evolution
inside a $15\,M_\odot$ star with a density profile
$\rho_\star\propto{}r^{-2.5}$ and a radius of $10^{11}$~cm. The jet is
injected at a radius $r_0=10^8$~cm with an opening angle
$\theta_0=20^\circ$, a Lorentz factor $\Gamma_0=1.5$ and a luminosity
$L_j=10^{51}$~erg~s$^{-1}$. Note that the scale of figures is enlarged
to emphasize the shape of the cocoon.}
\label{fig:sim1}}
\end{figure*}

To show some examples of the cocoon evolution we assume a star with
mass $M_\star=15\,M_\odot$, with radius $R_\star=10^{11}$~cm and
$\rho_\star\propto{}r^{-\alpha_\star}$. Figure~\ref{fig:sim1} shows
the cocoon evolution at fixed times (indicated in the panels). The
star has a power-law density profile $\rho_\star\propto{r}^{-2.5}$
between $10^8$ and $10^{11}$~cm. No funnel pre-evacuation is
assumed. The jet is injected with $L_j=10^{51}$~erg~s$^{-1}$ at
$r_0=10^8$~cm with $\theta_0=20^\circ$ and $\Gamma_0=1.5$. A
recollimation shock immediately collimates and slows down the jet
(upper left panel of Fig.~\ref{fig:sim1}). The jet emerges on the
stellar surface after $\sim5$~s. A total energy of
$1.5\times10^{51}$~erg has been dissipated to create the jet path into
the star. Of this energy, $\sim4\times10^{50}$~erg are stored into the
cocoon plasma, while $\sim10^{51}$~erg are given to the star through
the cocoon shock. Such energy is of the order of the explosion energy
of Ic SNe and enough to unbind the star. The jet reaches the star
surface with an opening angle of $\sim1^\circ$ and a Lorentz factor
$\Gamma_{\rm{br}}\simeq24$ and is therefore in causal contact.

\begin{figure*}
\includegraphics[width=.9\textwidth]{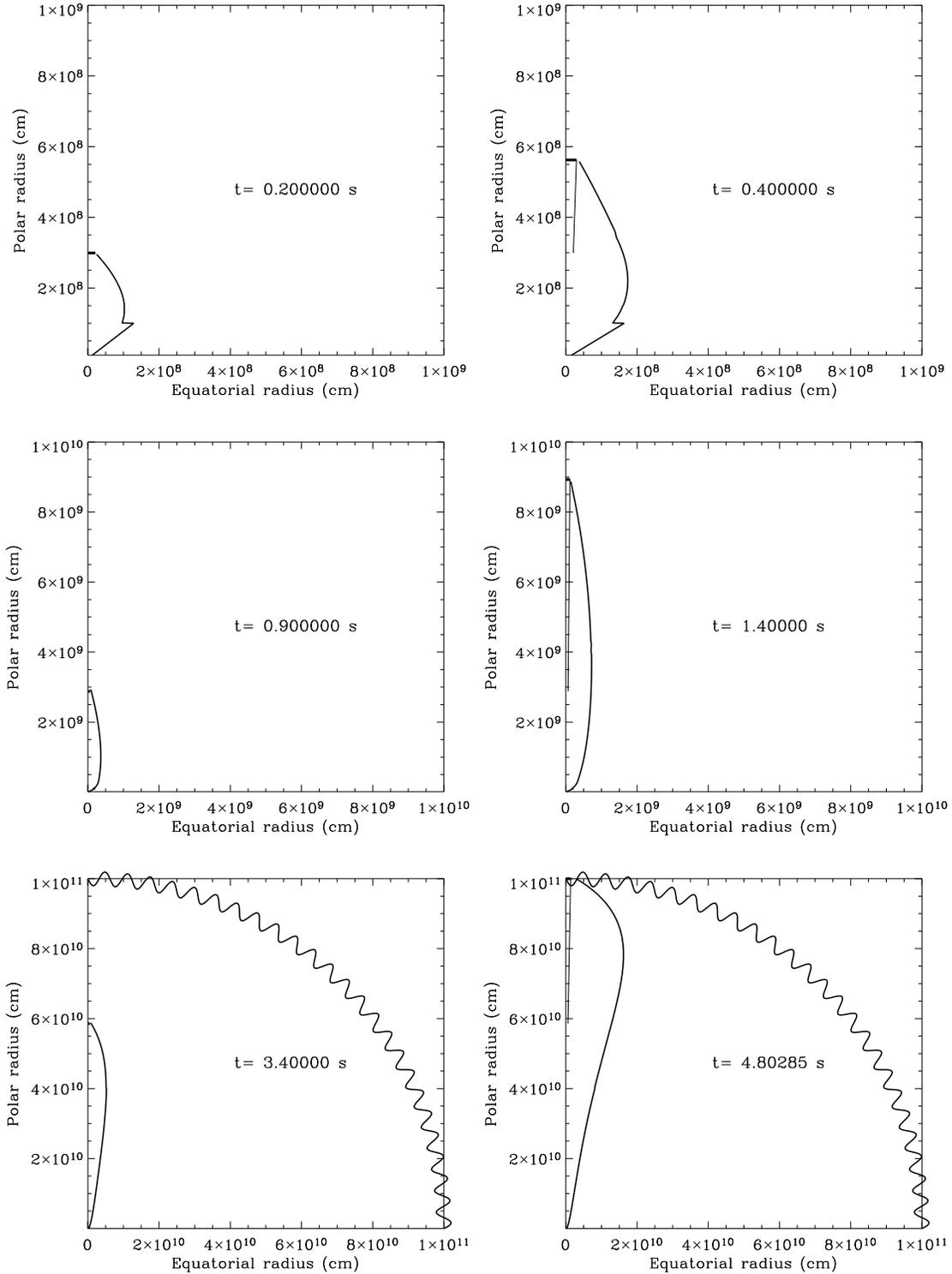}
\caption{{Same as Fig.~\ref{fig:sim1} but for a star with an
exponential cutoff in the density at large radii.}
\label{fig:sim2}}
\end{figure*}

Figure~\ref{fig:sim2} shows a similar simulation but where an
exponential cutoff of the stellar density of the form
$\exp[-r/(10^{10} {\rm{cm}})]$ has been added to simulate a more realistic
profile. The main effect of this change is on the shape of the cocoon
that opens as the exponential cutoff is reached. Also, the jet is
somewhat larger and has a larger Lorentz factor so that it turns out
to be only in marginal causal contact. These results compare
positively with analytic estimates~\cite{ref:laz05}.

\section{Precursors}

\begin{figure*}
\includegraphics[width=\textwidth]{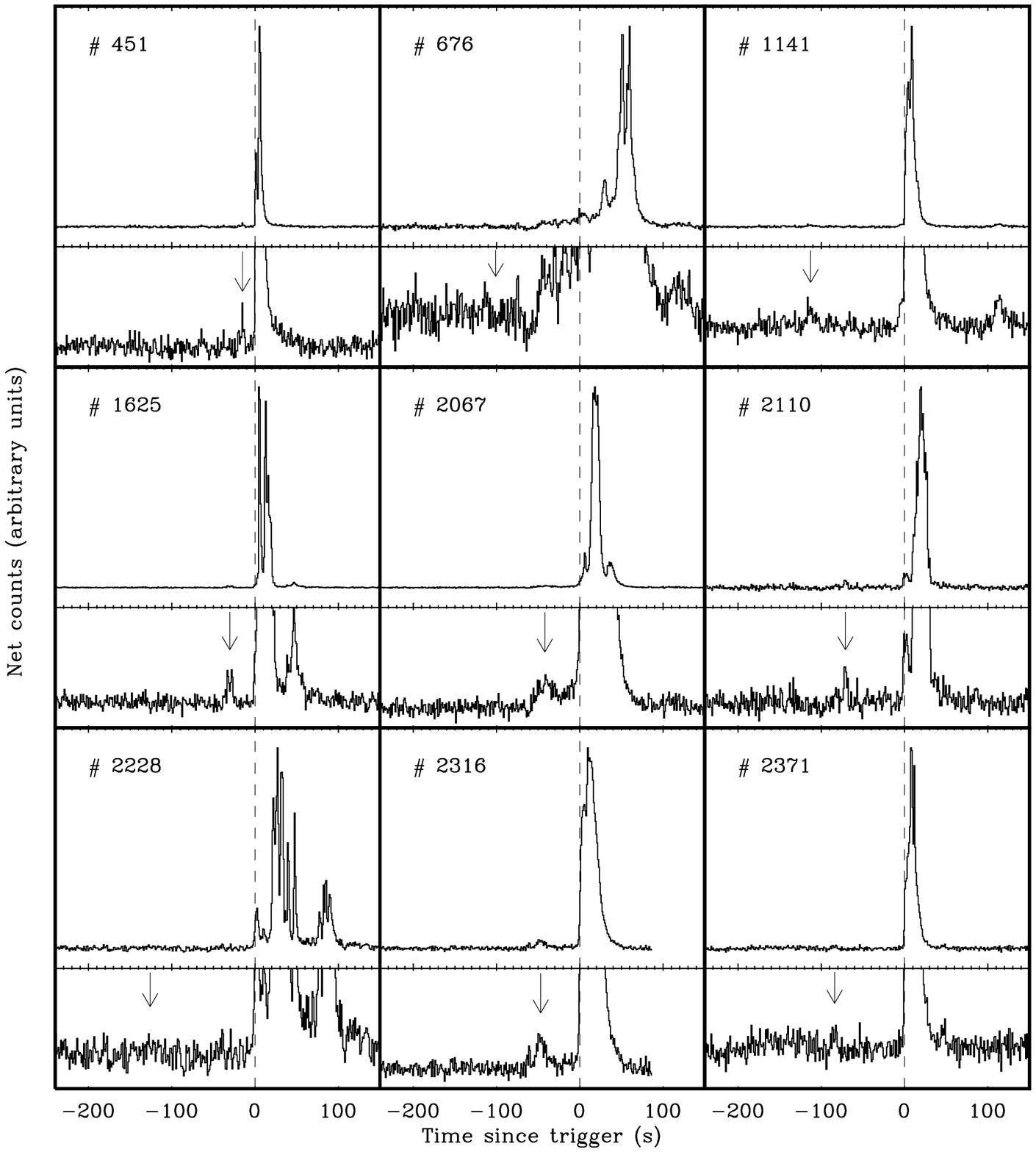}
\caption{Atlas of a fraction of the precursors detected in a sample of
bright long BATSE GRBs~\cite{ref:laz05b}.
\label{fig:precu}}
\end{figure*}

As the jet pierces the stellar surface, it simultaneously opens a
channel for the release of the cocoon. The cocoon material therefore
expands out of the star. The actual dynamics and radiative properties
of the cocoon are somewhat uncertain and several different predictions
have been made~\cite{ref:mes01,ref:ram02,ref:ram02b,ref:wax03}. What
is clear is that a total energy of several$\times10^{50}$ ergs is
released almost isotropically.

Observationally, associating a precursor with a transient event is
challenging, if at all possible. Attempts have been
made~\cite{ref:kos95,ref:laz05b} adopting conservative definitions,
and therefore only the tip of the iceberg was likely
discovered. Figure~\ref{fig:precu} shows a sub-sample of the
precursors discovered by Lazzati~\cite{ref:laz05b} in an analysis of
$\sim100$ bright long BATSE GRBs. He defined precursors as activity
that is detected before the GRB trigger from the same direction as the
GRB. The detected precursors are somewhat puzzling. A precursor
related to the jet release or to any event that takes place before the
$\gamma$-rays of the prompt phase are released is supposed to precede
the GRB by:
\begin{equation}
\Delta{T} = \frac{R_\gamma}{c\,\Gamma^2} \sim 0.03 \,
R_{\gamma,13}\,\Gamma_2^{-2} \qquad {\rm s}
\end{equation}
to be compared to the several tens of seconds of delays found
observationally. The only way\footnote{An alternative explanation
invoking the creation of a quark star has been recently put
forward~\cite{ref:pac05}.} to understand these precursors as related
to the cocoon breakout is to assume that the jet is initially dark in
$\gamma$-rays. We argue in the following that the jet spreads during
the $\sim100$~s of the energy release. If the observer is not lying on
the jet axis, there will be a time interval during which there will be
no $\gamma$-ray emission, even though the precursor had been
seen. Once the jet has spread, it enters the line of sight to the
observer and becomes visible, with a delay that depends on the
off-axis angle of the observer.

\section{Jet breakout and subsequent evolution}

As the jet reaches the stellar surface, it clears a channel for the
cocoon. The cocoon material is therefore now free to expand out of the
star and its pressure drops. We assume that from this moment on the
shock between the cocoon and the cold stellar material stalls, as a
consequence of the dropping cocoon pressure. This is equivalent to
assuming a constant volume of the cocoon cavity inside the star. The
pressure drop for the relativistic cocoon can be derived through
$dE_c=-\epsilon_c\,\Sigma_c\,c_s\,dt$ where $\Sigma_c$ is the area of
the free surface through which the cocoon material expands,
$\epsilon_c$ the cocoon energy density and $c_s=c/\sqrt{3}$ is the
sound speed of the relativistic gas.  Writing the cocoon volume as
$V_c\sim\Sigma_c\,r_\star$, we can obtain the pressure evolution:
\begin{equation}
p_c=p_{c,{\rm br}}\,\exp\left({-\frac{ct}{\sqrt{3}\,r_\star}}\right)
\end{equation}
where $p_{c,{\rm br}}$ is the cocoon pressure at the moment of shock
breakout.  As the cocoon pressure decreases, fresh jet material
passing through the cocoon is less tightly collimated. Under
isentropic conditions the jet Lorentz factor increases linearly with
the opening angle and pressure balance yields $\theta_j \propto
p_c^{-1/4}$, implying an exponentially increasing opening angle of the
form\footnote{Note that, if the jet is causally connected at breakout,
the jet would freely expand to an angle
$\theta_j=1/\Gamma_{j,{\rm{br}}}>\theta_{j,{\rm{br}}}$ of the order of
a few degrees. This would result in an initially constant opening
angle. The only effect on the final energy distribution of
eq.~\ref{eq:wow} is to increase the size of the jet core from
$\theta_{j,{\rm{br}}}$ to $1/\Gamma_{j,{\rm{br}}}$.}
\begin{equation}
\theta_j=\theta_{j, {\rm br}}\,\exp\left[\frac{c\,t}{4\sqrt{3}\,r_\star}
\right] .
\label{eq:thopen}
\end{equation}
Dissipative jet propagation gives similar results (with merely a
different numerical coefficient $\sim O(1)$ inside the exponential),
provided that $\Gamma$ varies roughly as a power of $\Sigma_j$.  

\section{Jet structure at large radii}

The angular distribution of integrated energy, as observed in the
afterglow phase, is computed by integrating the instantaneous
luminosity per unit solid angle from the moment the jet becomes
visible along a given line of sight ($t_{\rm{l.o.s.}}$) until the end
of the burst:
\begin{equation}
\frac{dE}{d\Omega}=\int_{t_{\rm{l.o.s.}}}^{T_{\rm{GRB}}}
\frac{dL}{d\Omega}\,dt \simeq\int_{t_{\rm{l.o.s.}}}^{T_{\rm{GRB}}}
\frac{L(t)}{\pi\,\theta_j^2(t)}\,dt .
\label{eq:struj0}
\end{equation}
$t_{\rm{l.o.s.}}$ is obtained by inverting eq.~\ref{eq:thopen}. Such
integration is valid provided that the jet opening angle at time
$T_{\rm{GRB}}$ is smaller than the natural opening angle of the jet:
$T_{\rm{GRB}}<4\sqrt{3}(r_\star/c)\log(\theta_0/\theta_{\rm{br}})$. For
the fiducial numbers assumed ($r_\star=10^{11}$~cm,
$\theta_{j,\rm{br}}=1^\circ$ and $\theta_0=30^\circ$) this corresponds
to $\sim100$~s comoving burst duration.  Assuming a jet with constant
luminosity $L$ and for all the lines of sight that satisfy $t_{\rm br}
< t_{\rm{l.o.s.}} \ll T_{\rm GRB}$, eq.~\ref{eq:struj0} gives the jet
structure
\begin{equation}
\frac{dE}{d\Omega} = \frac{2\sqrt{3}\,L\,r_\star}
{\pi\,c}\,\theta^{-2}  \qquad\theta_{j,{\rm{br}}}\le\theta\le\theta_0
\label{eq:wow}
\end{equation}
and $dE/d\Omega\sim$constant inside the core radius
$\theta_{j,{\rm{br}}}$.  This angular dependence, which characterizes
a ``structured jet'' or ``universal jet''
~\cite{ref:ros02,ref:zha02,ref:sal03,ref:lam05}, is of high
theoretical interest.  Jets with this beam pattern reproduce afterglow
observations. If the jet is powered by fall-back of material from the
star to the accretion disk, the mass accretion rate would be
anti-correlated with the radius of the star, for a given stellar
mass.

\begin{figure}[!t]
\includegraphics[width=\columnwidth]{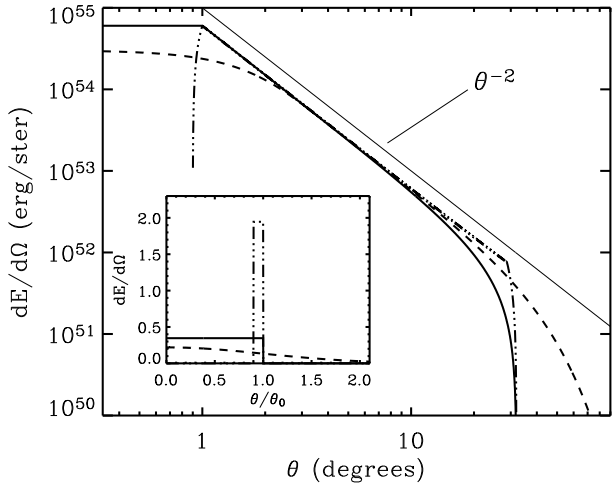}
\caption{{Energy distribution for the afterglow phase for three
instantaneous beam patterns (see inset). In all three cases a well
defined $dE/d\Omega\propto\theta^{-2}$ section is clearly
visible. Only the edge of the jet and its core show marginal
differences. The results shown are for a jet/star with
$L=10^{51}$~erg~s$^{-1}$, $T_{\rm{GRB}}=40$~s and
$r_\star=10^{11}$~cm. Inset: Instantaneous beam patterns that reach
the surface of the star. The solid line shows a uniform jet, dashed
line shows a Gaussian energy distribution, while the dash-dotted line
shows an edge brightened (or hollow) jet.}
\label{fig:pat}}
\end{figure}

In the above equations we have assumed for simplicity that the jet
reaching the surface of the star is uniform. As shown by
simulations~\cite{ref:zha03}, it is more likely that a Gaussian jet
emerges from the star. On the contrary, boundary layers may be
produced by the interaction of the jet with the collimating star,
resulting in edge brightened jets (see the inset of
Fig.~\ref{fig:pat}).  It can be shown easily that the $\theta^{-2}$
pattern does not depend on the assumption of
uniformity. Fig.~\ref{fig:pat} shows the integrated energy
distribution for uniform, Gaussian and hollow intrinsic jets. Even
though small differences are present at the edges (the jet core and
the outskirts), the general behavior is always
$dE/d\Omega\propto\theta^{-2}$.

\section{The $E_\gamma$-$\epsilon_{\rm{peak}}$ correlation}

\begin{figure}[!t]
\includegraphics[width=\columnwidth]{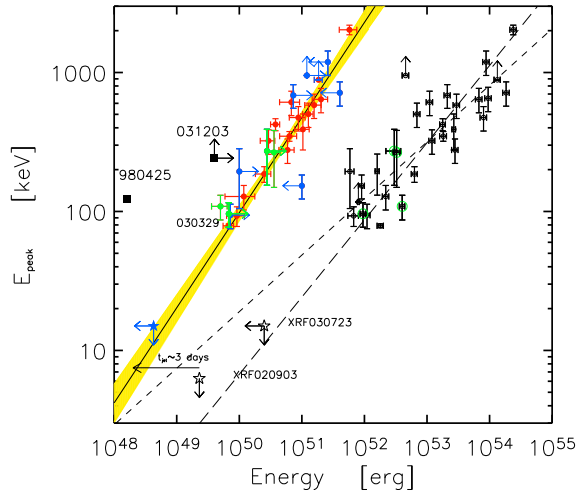}
\caption{{Rest frame peak energy $E_{peak}=E_{peak}^{obs}(1+z)$ versus
bolometric energy for the sample of GRBs with measured redshift. {\it
Filled circles}: isotropic energy corrected for the collimation angle
by the factor $(1-{\rm cos}\theta)$, for the events for which a jet
break in the light curve was observed. Grey symbols corresponds to
lower/upper limits.  The {\it Solid line} represents the best fit to
the correlation, i.e.  { $E_{peak} \sim 480 \, (E_\gamma/10^{51}{\rm
erg})^{0.7}$} keV.  {\it Open circles}: isotropic equivalent energy
$E_{\gamma,iso}$.  The {\it Dashed line} is the best fit to these
points.}
\label{fig:ghi}}
\end{figure}

Correlations between observed (and intrinsic) quantities are one of
the most remarkable results on the general properties of GRB
jets. They tell us that, despite the varied behavior of the observed
jet properties, they are not entirely random but follow a well defined
path. Among such correlations~\cite{ref:fra01,ref:pan01,ref:ama02} the
more recent and tightest is the so-called Ghirlanda
correlation~\cite{ref:ghi04} that finds a correspondence between the
beaming corrected $\gamma$-ray energy release of the prompt emission
and the typical photon frequency (the peak of the $\nu\,F(\nu)$
spectrum). This correlation is so tight that, within the present
accuracy, it is consistent with being an exact relation. The debate is
however alive on this issue~\cite{ref:fri05}.

The correlation is shown in Fig.~\ref{fig:ghi}. On the right of the
same graph the Amati~\cite{ref:ama02} correlation is shown for
comparison. The origin of these correlation is not understood so
far. It is believed that they may be due to a less dominant role of
synchrotron in the prompt radiative phase of the
burst~\cite{ref:ree05}.
 
\section{Standard candles and cosmography}

The Ghirlanda correlation is so tight that it can be used to
standardize the GRB candle~\cite{ref:ghi04a}. If one is able to
measure the break time of the GRB afterglow (and therefore constrain
the jet opening angle) and the typical frequency of photons in the
$\gamma$-ray spectrum, it is possible to predict the intrinsic energy
radiated in $\gamma$-rays in the prompt phase. This allows one to
measure the luminosity distance independently of redshift. If, in
addition, a redshift measurement is available, we can draw a Hubble
diagram and fit cosmological models to it.

\begin{figure}
\includegraphics[width=\columnwidth]{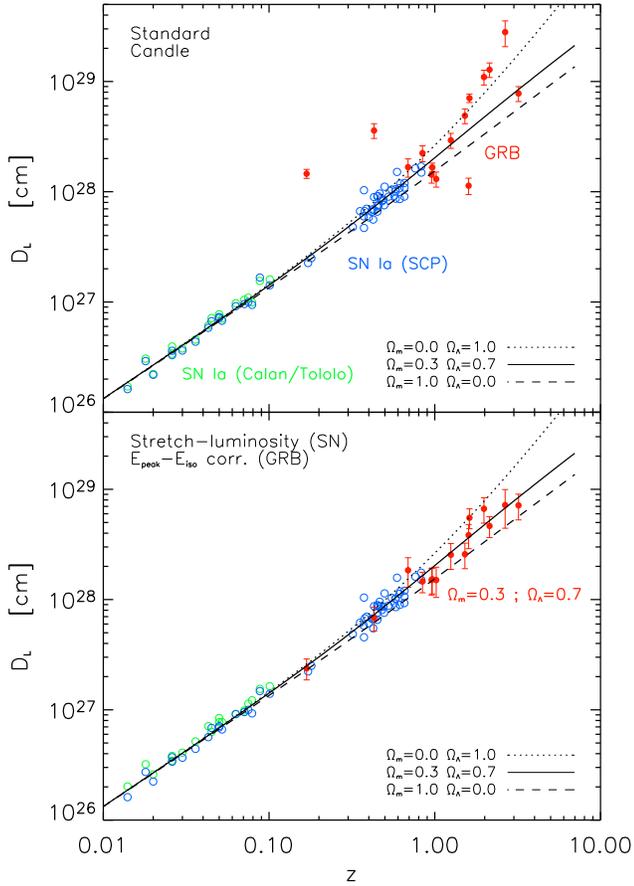}
\caption{{Classical Hubble diagram in the form of luminosity--distance
$D_L$ vs redshift $z$ for Supernova Ia and GRB. In the top panel the
SN Ia and GRBs are treated as standard candles (no corrections
applied); for GRBs $E_\gamma=10^{51}$ erg is assumed.  In the bottom
panel we have applied the stretching--luminosity and the
$E_\gamma$--$E_{peak}$ relations to SN Ia and GRBs, respectively. Note
that, for GRBs, the applied correction depends upon the specific
assumed cosmology, here for simplicity we assume the standard
$\Omega_M=0.3$, $\Omega_\Lambda=0.7$ cosmology.  Both panels also show
different $D_L(z)$ curves, as labeled.}
\label{fig:hub}}
\end{figure}

\begin{figure}
\includegraphics[width=\columnwidth]{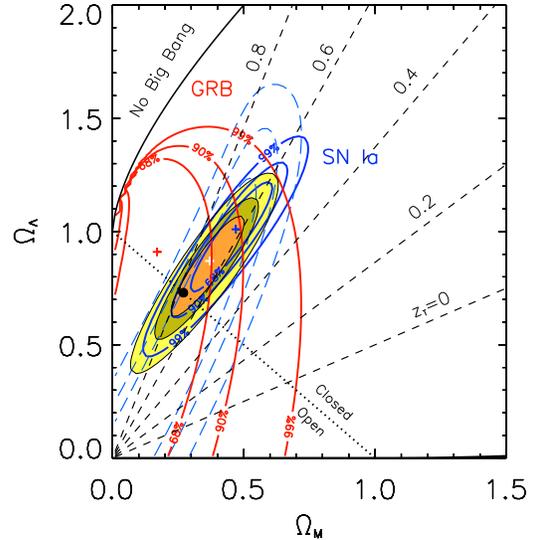}
\caption{{Constraints in the $\Omega_M-\Omega_\Lambda$ plane derived
from our GRB sample (15 objects, red contours), from the ``Gold" SN Ia
sample~\cite{ref:rie04} (156 objects; blue solid lines, derived
assuming a fixed value of $H_0=65$ km s$^{-1}$ Mpc$^{-1}$, and from
the subset of SNe Ia at $z>0.9$ (14 objects, blue long dashed lines).
The three colored ellipses are the confidence regions (orange: 68\%;
light green: 90\%; yellow: 99\%) for the combined fit of type Ia
SN+GRB samples.  Dashed lines correspond to the changing sign of the
cosmic acceleration [i.e. $q(z)=0$] at different redshifts, as
labeled.  Crosses are the centers of the corresponding contours (red:
GRBs; blue: SNe Ia, white: GRB+SN Ia).  The black dot marks the
$\Lambda$CDM cosmology.  The dotted line corresponds to the
statefinder $r=1$, in this case it coincides with the flatness
condition.}
\label{fig:cosmo}}
\end{figure}

Figure~\ref{fig:hub} shows the GRBs and SNe Hubble diagram before and
after the empirical corrections (stretch-luminosity for SNe and
Ghirlanda for GRBs). Figure~\ref{fig:cosmo} shows instead the
constraints that can be obtained in the $\Omega_M-\Omega_\Lambda$
plane with SNe, GRBs and a simultaneous fit of the two
samples~\cite{ref:fir05}. Even though the simultaneous fit is clearly
dominated by SNe, the perspective for GRBs, should the Ghirlanda
correlation be confirmed, are very good. Their appeal is, more than to
give an independent confirmation of the SNe result, to allow for an
expansion to high redshift of the universe that can be sampled. This
is an ideal set-up to investigate on the possible evolution of the
properties of $\Omega_\Lambda$ to understand better the origin of dark
energy.

\begin{figure}[!t]
\includegraphics[width=\columnwidth]{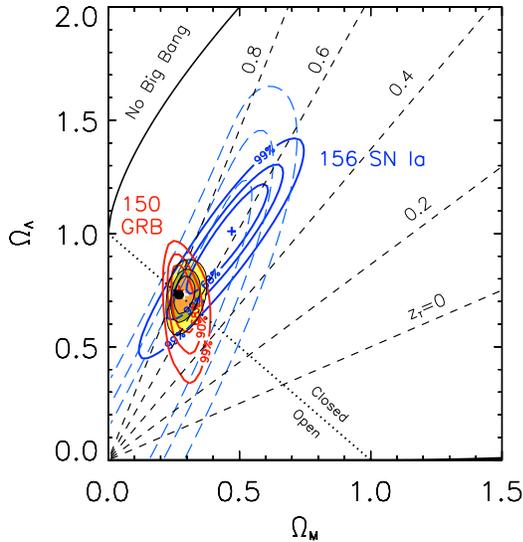}
\caption{{Swift simulated $\Omega_M-\Omega_\Lambda$ plane.}
\label{fig:swi}}
\end{figure}

With the launch of Swift, it is envisaged that the sample of GRBs with
all the necessary data will soon become comparable to the SN
sample. Fig.~\ref{fig:swi} shows a prediction of how the cosmological
constraints from GRBs in the Swift era may look.

\section{Discussion}

We have discussed the properties of GRB jets within the framework of
the GRB-SN association. We showed that the interaction of the jet with
the star produces a cocoon that modifies the jet properties. Even
though the jet may be released in a stationary process, with constant
properties in time, the interaction with the star and the cocoon
creates a time-dependent opening angle of the jet. In the prompt phase
the main consequence is that the beginning of the $\gamma$-ray
emission depends on the observer. In the afterglow phase, when the
time scales of the prompt emission are irrelevant, the jet is seen as
a structured outflow with a well defined profile. This, with some
additional constraints on the luminosity and duration of the energy
injection, can create a universal jet profile~\cite{ref:ros02}.

Observationally, the properties of GRB jets are not independent. In
particular, the Ghirlanda correlation~\cite{ref:ghi04} tells us that
if the jet opening angle (or viewing angle) and the total energy
output are known, the typical frequency of photons can be
predicted. The correlation is so tight that can be used, analogously
to the stretch-luminosity correlation in type Ia SN~\cite{ref:phi93},
to standardize the GRB candle. GRBs can therefore be used as high
redshift standard candles to constrain any possible evolution in the
properties of the dark energy.

\section*{Acknowledgments}
This research has been partly supported by the NSF grant AST-0307507


\begin{thebibliography}{99}
\bibitem{ref:wos93} Woosley S.~E., 1993, ApJ, 405, 273
\bibitem{ref:pac98} Paczy\'nski B., 1998, ApJ, 494, L45 
\bibitem{ref:gal98} Galama T.~J., et al., 1998, Nature, 395, 670
\bibitem{ref:zeh04} Zeh A., Klose S., Hartmann D.~H., 2004, ApJ, 609,
  952
\bibitem{ref:van04} Vanderspek R., et al., 2004, ApJ, 617, 1251 
\bibitem{ref:sta03} Stanek K.~Z., et al., 2003, ApJ, 591, L17
\bibitem{ref:hjo03} Hjorth J., et al., 2003, Nature, 423, 847 
\bibitem{ref:mat03} Matheson T., et al., 2003, ApJ, 599, 394 
\bibitem{ref:mal04} Malesani D., et al., 2004, ApJ, 609, L5
\bibitem{ref:fra01} Frail D.~A., et al., 2001, ApJ, 562, L55
\bibitem{ref:pan01} Panaitescu A., Kumar P., 2001, ApJ, 554, 
  667 
\bibitem{ref:ros02} Rossi E., Lazzati D., Rees M.~J., 2002, MNRAS,
  332, 945
\bibitem{ref:ama02} Amati L., et al., 2002, A\&A, 390, 81
\bibitem{ref:ghi04} Ghirlanda G., Ghisellini G., Lazzati D., 2004,
  ApJ, 616, 331
\bibitem{ref:laz05} Lazzati D., Begelman M. C., 2005, ApJ
  subm. (astro-ph/0502084)
\bibitem{ref:ghi04a} Ghirlanda G., Ghisellini G., Lazzati D., 
  Firmani C., 2004, ApJ, 613, L13
\bibitem{ref:mar94} Marti J.~M., Mueller E., Ibanez J.~M., 
  1994, A\&A, 281, L9
\bibitem{ref:matz03} Matzner C.~D., 2003, MNRAS, 345, 575 
\bibitem{ref:kom60} Kompaneets A.~S., 1960, Soviet Phys. Dokl., 5, 46
\bibitem{ref:mes01} M{\' e}sz{\' a}ros P., Rees M.~J., 2001, 
  ApJ, 556, L37
\bibitem{ref:ram02} Ramirez-Ruiz E., MacFadyen A.~I., 
  Lazzati D., 2002, MNRAS, 331, 197
\bibitem{ref:ram02b} Ramirez-Ruiz E., Celotti A., Rees M.~J., 
  2002, MNRAS, 337, 1349
\bibitem{ref:wax03} Waxman E., M{\' e}sz{\' a}ros P., 2003, 
  ApJ, 584, 390
\bibitem{ref:kos95} Koshut T.~M., Kouveliotou C., Paciesas W.~S., 
  van Paradijs J., Pendleton G.~N., Briggs M.~S., Fishman G.~J., 
  Meegan C.~A., 1995, ApJ, 452, 145
\bibitem{ref:laz05b} Lazzati D., 2005, MNRAS, 357, 722
\bibitem{ref:pac05} Paczy\'nski B, Haensel P., 2005, MNRAS submitted
  (astro-ph/0502297)
\bibitem{ref:zha02} Zhang B., M{\' e}sz{\' a}ros P., 2002, 
  ApJ, 571, 876 
\bibitem{ref:sal03} Salmonson J.~D., 2003, ApJ, 592, 1002 
\bibitem{ref:lam05} Lamb D.~Q., Donaghy T.~Q., Graziani 
  C., 2005, ApJ, 620, 355
\bibitem{ref:zha03} Zhang W., Woosley S.~E., MacFadyen 
  A.~I., 2003, ApJ, 586, 356
\bibitem{ref:fri05} Friedman A. S., Bloom J. S., 2005, ApJ submitted
  (astro-ph/0408413)
\bibitem{ref:ree05} Rees M. J., Meszaros P., 2005, ApJ submitted
  (astro-ph/0412702)
\bibitem{ref:fir05} Firmani C., Ghisellini G., Ghirlanda G.,
  Avila-Reese V., 2005, MNRAS in press (astro-ph/0501395)
\bibitem{ref:rie04} Riess A.~G., et al., 2004, ApJ, 607, 665
\bibitem{ref:phi93} Phillips M.~M., 1993, ApJ, 413, L105
\end{thebibliography}
\end{document}